\documentclass[12pt]{article}
\usepackage[english,russian]{babel}
\usepackage{epsfig}
\begin{document}
\large
\par
\begin{center}
{\bf Examination of unitarity condition (positive definiteness of
expression for transition probabilities) at three neutrino
oscillations in vacuum}
\par
\vspace{0.3cm} Beshtoev Kh. M.
\par
\vspace{0.3cm} Joint Institute for Nuclear Research, Joliot Curie
6, 141980 Dubna, Moscow region Russia.
\end{center}
\vspace{0.3cm}

\par
Abstract

\par
This work has shown that at strict fulfilment of condition $\Delta
m^2_{1 3} = \Delta m^2_{1 2} + \Delta m^2_{2 3}$ the expression
for probability of $\nu_e \to \nu_e$ transitions $P_{\nu_e \to
\nu_e}(t)$ is positively defined at every values of $\theta$ and
$\beta$ while at any arbitrarily small deviation from this
condition it becomes negative. In order to make this expression
for probability transitions positively defined, it is necessary to
put a limitation on angle mixing $\beta$ at fixed value of
$\theta= 32.45^o$ (i.e. the value for $\beta$ must be $\beta \le
15^o \div 17^o$).

\section{Introduction}

\par
The suggestion that in analogy with $K^{o},\bar K^{o}$
oscillations, there could be neutrino-antineutrino oscillations (
$\nu \rightarrow \bar \nu$), was considered by Pontecorvo [1] in
1957. It was subsequently considered by Maki et al. [2] and
Pontecorvo [3] that there could be mixings (and oscillations) of
neutrinos of different flavors (i.e., $\nu _{e} \rightarrow \nu
_{\mu }$ transitions).
\par
In the general case there can be two schemes (types) of neutrino
mixings (oscillations): mass mixing schemes and charge  mixings
scheme (as it takes place in the vector dominance model or vector
boson mixings in the standard model of electroweak interactions)
[4].
\par
In the scheme of charge mixings the oscillation parameters are
expressed through weak interaction couple constants (charges) and
neutrino masses [4].
\par
In the both cases the neutrino mixing matrix $V$ can be given [4]
in the following convenient form proposed by Maiani [6]($\theta =
\theta_{12}, \beta = \theta_{13}, \gamma = \theta_{23}$):
\par
$$
{V {=} \left( \begin{array} {ccc}1& 0 & 0 \\
0 & c_{\gamma} & s_{\gamma} \\ 0 & -s_{\gamma} & c_{\gamma} \\
\end{array} \right)\!\! \left( \begin{array}{ccc} c_{\beta} & 0 &
s_{\beta} \exp(-i\delta) \\ 0 & 1 & 0 \\ -s_{\beta} \exp(i\delta)
& 0 & c_{\beta} \end{array} \right)\!\! \left( \begin{array}{ccc}
c_{\theta} & s_{\theta} & 0 \\ -s_{\theta} & c_{\theta} & 0 \\ 0 &
0 & 1 \end{array}\right)} , \eqno(1)
$$
where
$$
c_{e \mu} = \cos {\theta } , \quad s_{e \mu} =\sin{\theta}, \quad
c^2_{e \mu} + s^2_{e \mu} = 1 ;
$$
$$
c_{e \tau} = \cos {\beta }, \quad s_{e \tau} =\sin{\beta}, \quad
c^2_{e \tau} + s^2_{e \tau} = 1 ;
$$
$$
c_{\mu \tau} = \cos {\gamma} , \quad s_{\mu \tau} =\sin{\gamma},
\quad c^2_{\mu \tau} + s^2_{\mu \tau} = 1 ;
$$
$$
 \exp(i\delta) = \cos{\delta } + i \sin{\delta} .
$$

\par
Using the above matrix $V$, we can connect the wave functions of
physical neutrino states $\Psi_{\nu_e}, \Psi_{\nu_\mu},
\Psi_{\nu_\tau}$ with the wave functions of intermediate neutrino
states $\Psi_{\nu_1}, \Psi_{\nu_2}, \Psi_{\nu_3}$ and write it
down in a component-wise form [5]:
\par
$$
\Psi_{\nu_l} = \sum^{3}_{k=1}V^{*}_{\nu_l \nu_k} \Psi_{\nu_{k}},
$$
$$
\Psi_{\nu_k} = \sum^{3}_{k=l}V_{\nu_k \nu_l} \Psi_{\nu_{l}},
\qquad l = e, \mu, \tau , \qquad k = 1 \div 3 , \eqno(2)
$$
where $\Psi_{\nu_{k}}$ is a wave function of neutrino with
momentum $p$ and mass $m_{k}$. We suppose that neutrino mixings
(oscillations) are virtual if neutrinos have different masses (if
we suppose that these transitions are real, as it is supposed in
the standard theory of neutrino oscillations, then it is necessary
to accept that expression (2) is based on a supposition that
masses difference of $\nu_{k}$ neutrinos is so small that coherent
neutrino states are formed in the weak interactions (computation
has shown that this condition is not fulfilled, i.e. neutrino as
wave packet is unstable and decays)).
\par
$$
\Psi_{\nu_{k}}(t) = e^{-i E_k t} \Psi_{\nu_{k}}(0) . \eqno(3)
$$
\par
Then
$$
\Psi_{\nu _{l}(t)} =\sum^{3}_{k=1} e^{-i E_k t} V^{*}_{\nu_l
\nu_k} \Psi_{\nu_{k}}(0)  . \eqno(4)
$$
Using unitarity of matrix $V$ or expression (2) we can rewrite
expression (4) in the following form:
\par
$$
\Psi_{\nu _{l}}(t) = \sum^{}_{l'= e,\mu, \tau} \sum^{3}_{k=1}
V_{\nu_{l'} \nu_k} e^{-i E_k t} V^{*}_{\nu_l \nu_k}
\Psi_{\nu_{l'}(0)} , \eqno(5)
$$
and introducing symbol $b_{\nu_{l}\nu _{l'}}(t)$
$$
b_{\nu_{l}\nu _{l'}}(t) = \sum^{3}_{k=1} V_{\nu_{l'} \nu_k} e^{-i
E_k t} V^{*}_{\nu_l \nu_k} , \eqno(6)
$$
we obtain
$$
\Psi_{\nu_{l}}(t) = \sum^{}_{l'=e, \mu, \tau}
b_{\nu_{l}\nu_{l'}}(t) \Psi_{\nu_{l'}}(0) , \eqno(7)
$$
where $b_{\nu_{l} \nu_{l'}}(t)$-is the amplitude of transition
probability $\Psi_{\nu_{l}}  \rightarrow  \Psi_{\nu_{l'}}$.
\par
\noindent And the corresponding expression for transition
probability $\Psi_{\nu_{l}} \rightarrow \Psi_{\nu_{l'}}$ is:
\par
$$
P_{\nu_{l}\nu_{l'}}(t) =\mid \sum^{3}_{k=1} V_{\nu_l' \nu_k} e^{-i
E_k t} V^{*}_{\nu_l \nu_k} \mid^{2} . \eqno(8)
$$
Using expression (8) in work [6] the amplitudes and expression for
probability of three neutrino transitions (oscillations) for all
interesting cases (i.e., for $\nu_e \to \nu_e, \nu_\mu, \nu_\tau$,
$\nu_\mu \to \nu_e, \nu_\mu, \nu_\tau$, $\nu_\tau \to \nu_e,
\nu_\mu, \nu_\tau$). In these works was done examination is the
expression for probability of $\nu_e \leftrightarrow \nu_e$
neutrino transitions $P_{\nu_e \nu_e}$
$$
P_{\nu_e \to \nu_e} (t)= 1 - cos^4(\beta)sin^2(2 \theta) sin^2(- t
(E_1-E_2)/2) -
$$
$$
cos^2(\theta) sin^2(2 \beta) sin^2(- t (E_1-E_3)/2) - \eqno(9)
$$
$$
-sin^2(\theta) sin^2(2 \beta) sin^2(- t (E_2-E_3)/2) .
$$
positive definite values at all values for $\beta, \theta, E_1,
E_2, E_3$ and $t$? It was shown that at arbitrary values of the
parameters this expression for probability is not a positive
definite value. Then to make this expression for probability
positively defined, it is necessary to put limitation on values of
these parameters.
\par
This work continues the study of this question.

\section{Examination of positive definiteness of the expression for probability
$P_{\nu_e \to \nu_e} (t)$ transitions}

\par
The aim of this work is to examine positive definiteness of the
expression for probability $P_{\nu_e \to \nu_e} (t)$ transitions
expr. (9). For this purpose we will fulfill graphical modelling of
this function. This expression contains 7 parameters - $\theta$,
$\beta$, $\Delta m^2_{1 2}$, $\Delta m^2_{2 3}$, $\Delta
m^2_{13}$, $t$, $E_{\nu_e} \simeq p_{\nu_e} c$. There is one
connection between neutrino masses $\Delta m^2_{1 3} = \Delta
m^2_{1 2} + \Delta m^2_{2 3}$ therefore this expression contain
only 6 independent parameters. Obtaining the extremums of this
expression in the general case is a serious problem. Instead of it
we decided to simplify this problem and used the following values
for parameters obtained in the experiments:
$$
sin^2 (2 \theta_{\nu_e \nu_\mu}) \cong 0.83 , \quad \theta =
32.45^o , \quad \Delta m^2_{1 2} = 8.3 \cdot 10^{-5} {eV}^2,
\eqno(10)
$$
in [7] and
$$
sin^2 (2\gamma_{\nu_\mu \nu_\tau})  \cong  1, \quad \gamma \cong
\frac{\pi}{4} \quad \Delta m^2_{2 3} = 2.5 \cdot 10^{-3} {eV}^2,
\eqno(11)
$$
in [8]. Value for $E_{\nu_e}$ is free. We expressed all lengths of
oscillations through the length of oscillations determined by
$\nu_1, \nu_2$ masses. The values of $\Delta m^2_{1 3} = \Delta
m^2_{1 2} + \Delta m^2_{2 3}$ are $2.583 \cdot 10^{-3} {eV}^2$,
and $2.417 \cdot 10^{-3} {eV}^2$.

Thus in expression (9) there is remained dependence only of two
parameters $\beta$ and $t$ and then this expression for the first
case $\Delta m^2_{1 3} = 2.583 \cdot 10^{-3} {eV}^2$ can be
rewritten in the following form:
$$
P_{\nu_e \to \nu_e} (t)= 1 - cos^4(\beta)sin^2(2 \theta)
sin^2(R/L_{1 2}) -
$$
$$
- cos^2(\theta) sin^2(2 \beta) sin^2(30.120 R/L_{1 2}) - \eqno(12)
$$
$$
-sin^2(\theta) sin^2(2 \beta) sin^2(31.120 R/L_{1 2}) .
$$
where $L_{1 2} = 1.27 \frac{c p_{\nu_e}}{\Delta m^2_{1 2}}$ is the
length of oscillations determined by $\nu_1, \nu_2$ neutrino
masses and then lengths of oscillations determined by $\nu_1,
\nu_3$ and $\nu_2, \nu_3$ neutrino masses are expressed through
$L_{1 2}$ oscillations length of $\nu_1, \nu_2$ neutrinos ($2.5
10^{-3}/8.3 10^{-5} = 30.120$ $2.583 10^{-3}/8.3 10^{-5} =
31.120$) and $R$ is a distance from the neutrino source.
\par
We fulfilled modelling of this expression for $\beta = 10^o$,
$15^o$, $20^o$, $25^o$, $30^o$, $35^o$, $40^o$, $45^o$, $50^o$,
$55^o$ for $t = 0 \div 4 \pi$ for the two above mentioned cases
when $\Delta m^2_{1 3} = 2.583 \cdot 10^{-3} {eV}^2$, and $2.417
\cdot 10^{-3} {eV}^2$. The check has confirmed that for the all
above cases the unitarity condition is fulfilled (i.e. $P_{\nu_e
\to \nu_e}(t) \geq 0$)
\par
Pictures 1, 2, 3, 4 show modelling of probability of $P_{\nu_e \to
\nu_e}(t)$ for $\beta = 15^o, 25^o, 35^o, 45^o$ at $\Delta m^2_{1
3} = 2.583 \cdot 10^{-3} {eV}^2$ . From this pictures it is seen
that the unitarity condition is fulfilled when we take precision
values (relations) for lengths of oscillations (i.e. the condition
$\Delta m^2_{1 3} = \Delta m^2_{1 2} + \Delta m^2_{2 3}$ is
strongly fulfilled). At an arbitrary small deviation from the
precision values (relations) for lengths of oscillations the
unitarity condition is violated as it is well seen in Figure 5
($\beta = 45^o$, $31.120 \to 31.320$) and then to fulfil this
condition, it is necessary to apply the limitation on angle
$\beta$.

\newpage
\begin{figure}[h!]
   \begin{center}
   \includegraphics[width=10cm]{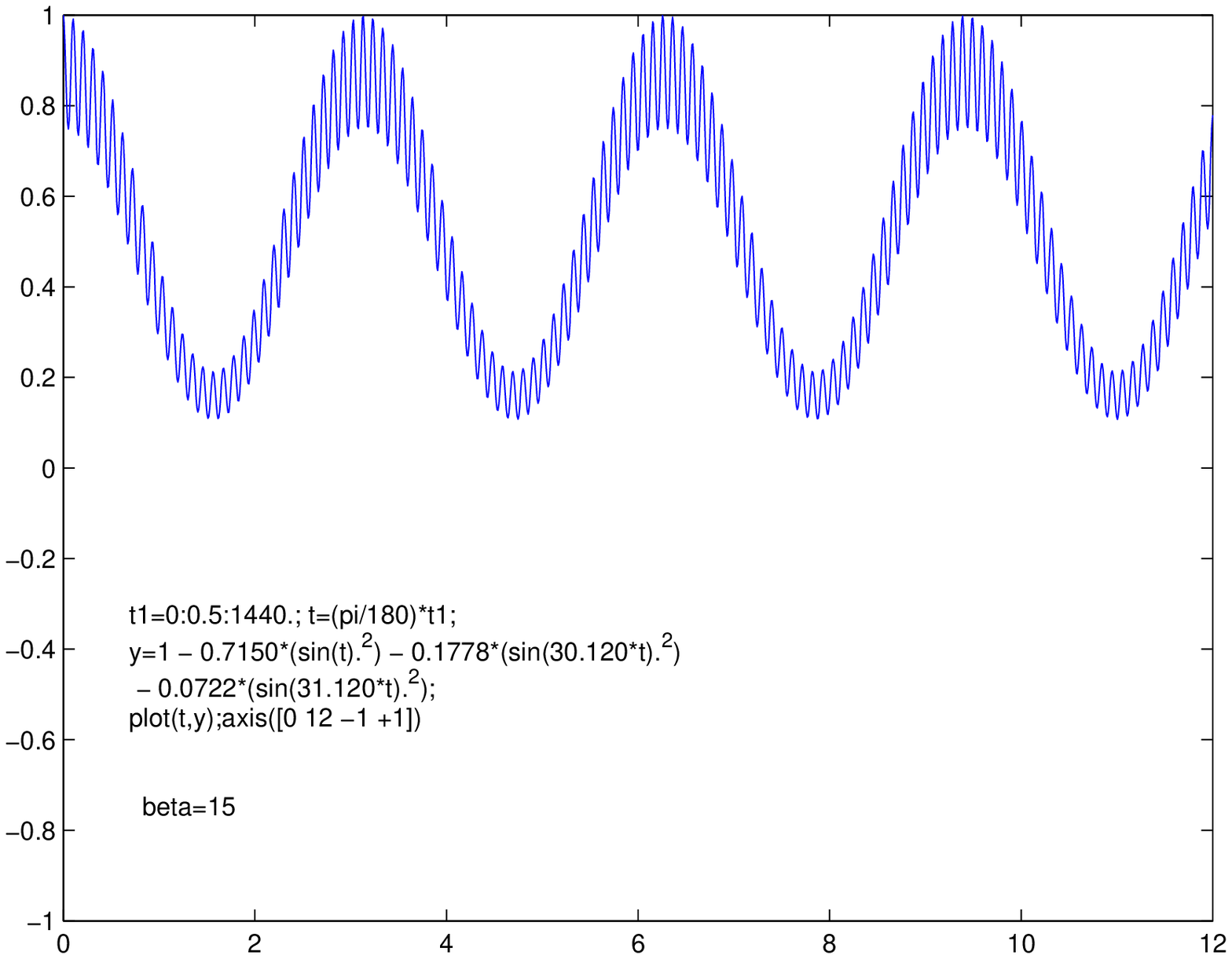}
   \end{center}
   \end{figure}

{\sf Figure~1:}~ $P_{\nu_e \to \nu_e}(t)$ at $\theta = 32.45^o,
\beta = 15^o$. \\

\begin{figure}[h!]
   \begin{center}
   \includegraphics[width=10cm]{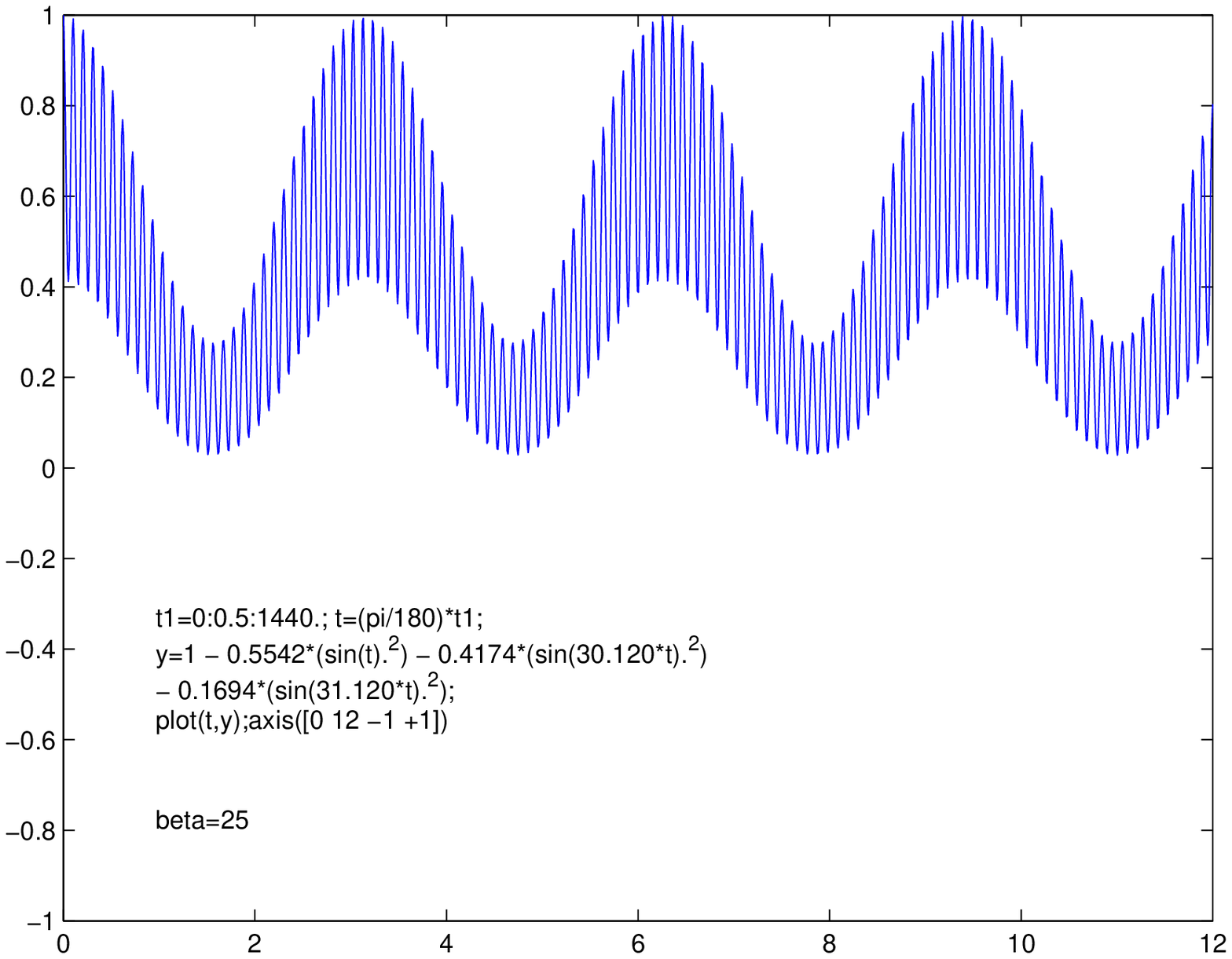}
   \end{center}
   \end{figure}

{\sf Figure~2:}~ $P_{\nu_e \to \nu_e}(t)$ at $\theta = 32.45^o,
\beta = 25^o$. \\

\newpage
\begin{figure}[h!]
   \begin{center}
   \includegraphics[width=10cm]{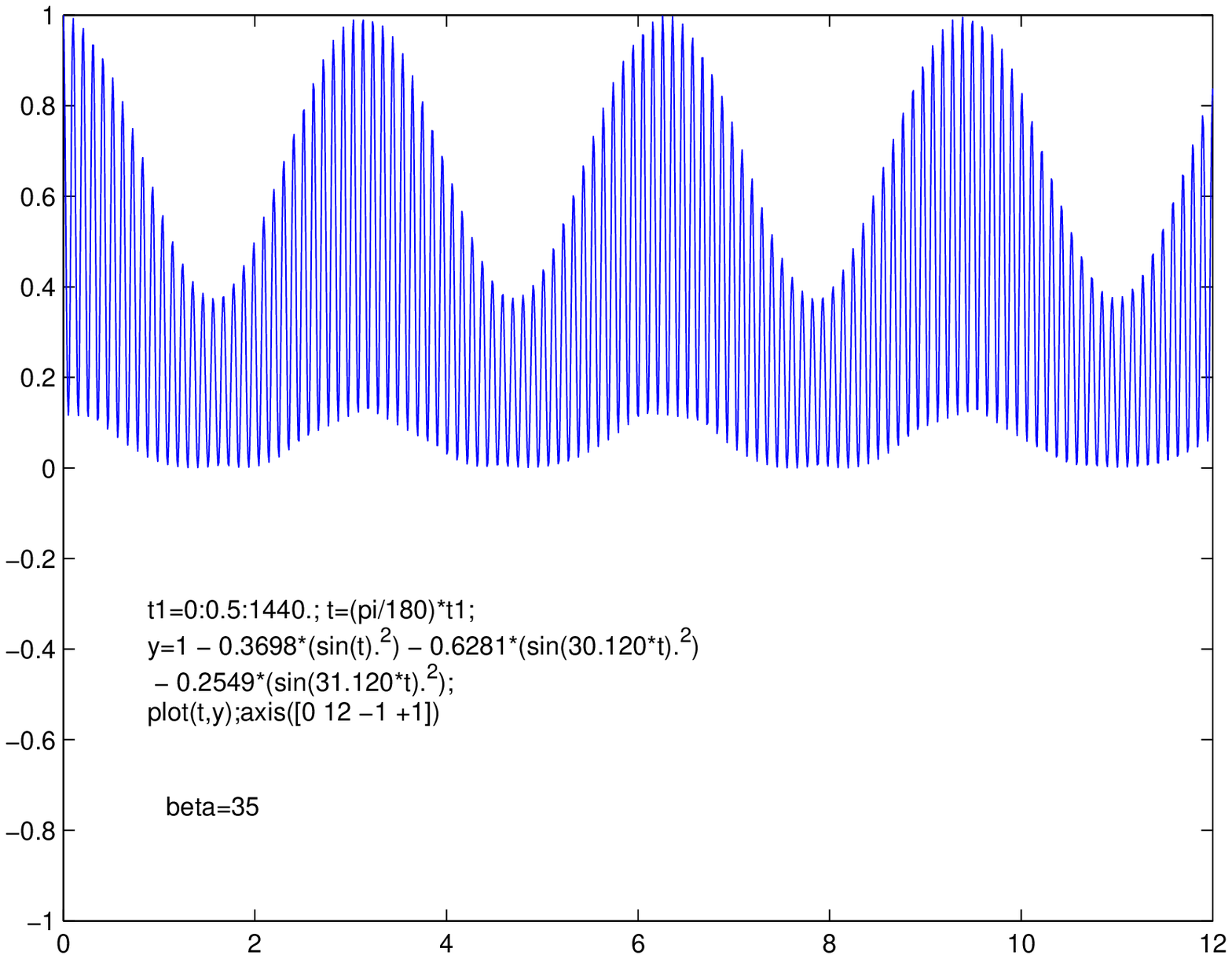}
   \end{center}
   \end{figure}

{\sf Figure~3:}~ $P_{\nu_e \to \nu_e}(t)$ at $\theta = 32.45^o,
\beta = 35^o$. \\

\begin{figure}[h!]
   \begin{center}
   \includegraphics[width=10cm]{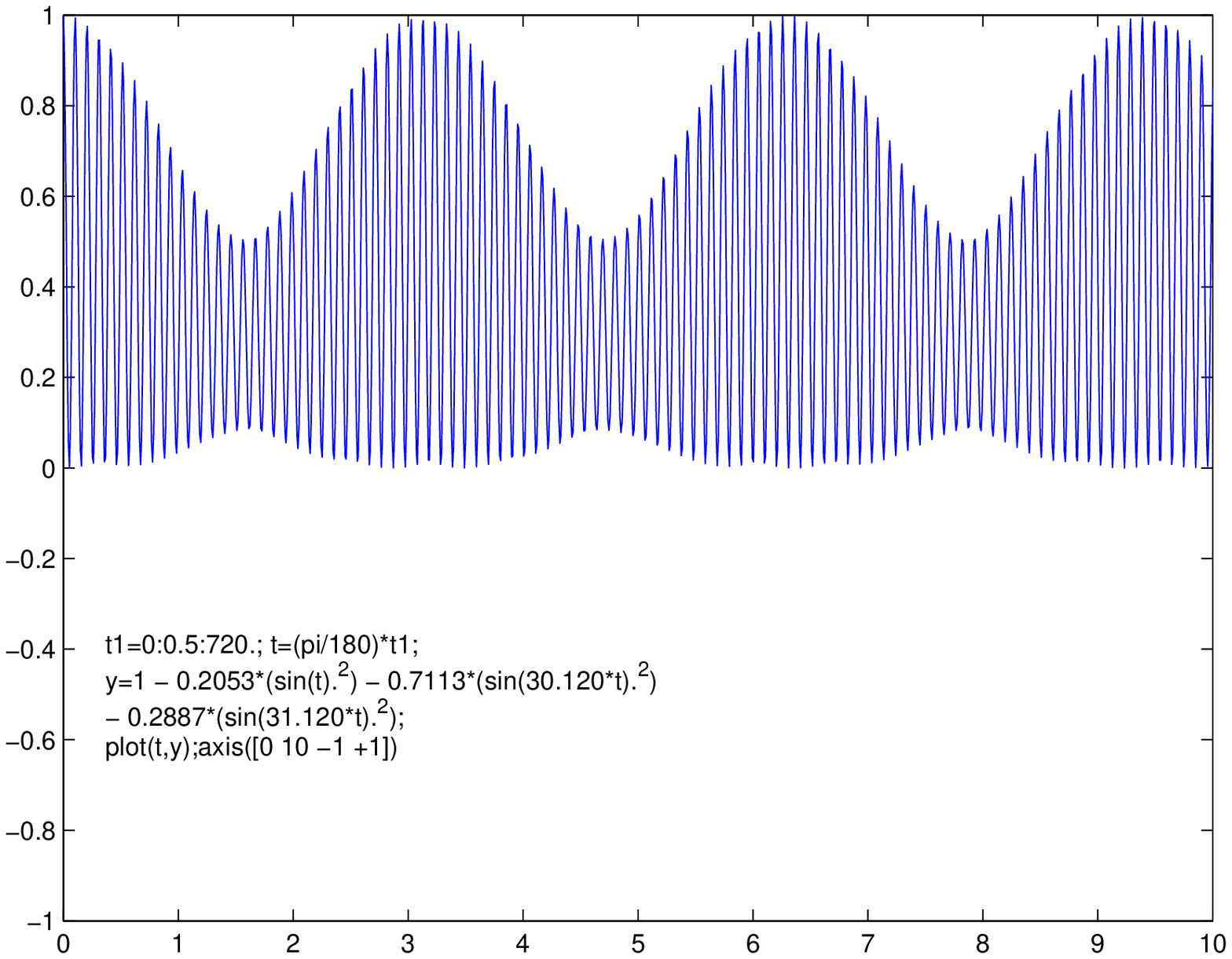}
   \end{center}
   \end{figure}

{\sf Figure~4:}~ $P_{\nu_e \to \nu_e}(t)$ at $\theta = 32.45^o,
\beta
= 45^o$. \\

\newpage
\begin{figure}[h!]
   \begin{center}
   \includegraphics[width=10cm]{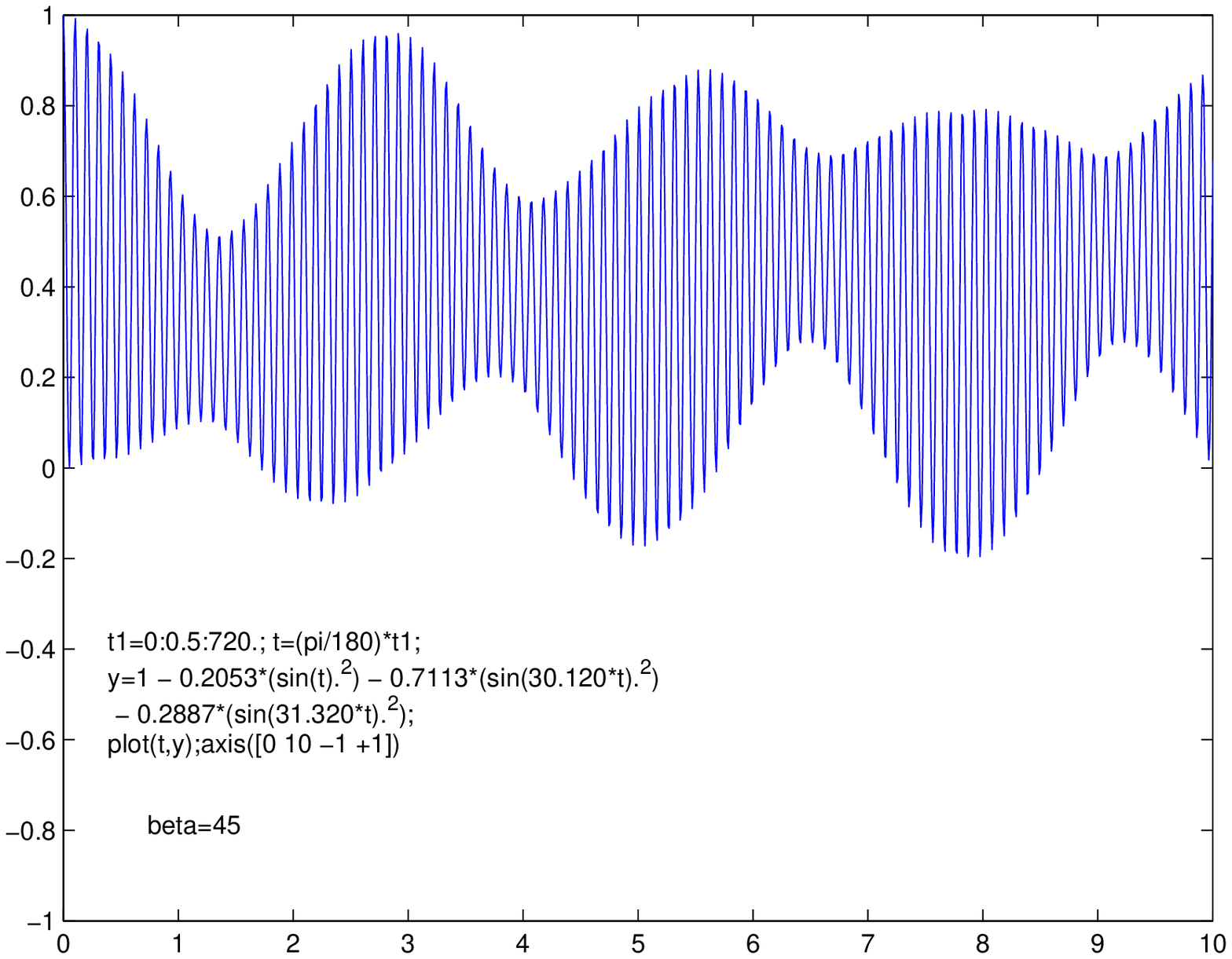}
   \end{center}
   \end{figure}

{\sf Figure~5:}~ $P_{\nu_e \to \nu_e}(t)$ at $\theta = 32.45^o,
\beta
= 45^o$, ($31.120 \to 31.320$) \\

\par
In the previous work [9] there was graphical modelling of this
function (9) by using the following values for [11] $\theta =
32.45^o, \Delta m^2_{1 2} = 8.3\cdot 10^{-5} eV^2$ and for [12] $
\Delta m^2_{2 3} = 2.5 \cdot 10^{-3} eV^2$, for the cases when
$\Delta m^2_{13} = 10^{-5} eV^2 \quad (L_{1 3}/L_{1 2} = 0.12),
\quad 5.7\cdot 10^{-5} eV^2 \quad (L_{1 3}/L_{1 2} = 0.69),\quad
8.3\cdot 10^{-4} eV^2 \quad (L_{1 3}/L_{1 2} = 10)$ (for checking)
for different values of $\beta = 10^o \div 45^o$ and established
that the value for $P_{\nu_e \to \nu_e} (t)$ become a positively
defined value at $\beta \leq 15^o \div 17^o$ ($P_{\nu_e \to \nu_e}
(t) \simeq 0$ at some values of $t$). Figures 6, 7, 8, 9, 10, 11
show results of modelling for $\beta = 10^o$, $\beta = 30^o$ for
the above indicated three cases. From these figures we see that at
$\beta = 10^o$ the unitarity condition is fulfilled while this
condition is violated at $\beta = 30^o$.

\newpage
\begin{figure}[h!]
   \begin{center}
   \includegraphics[width=10cm]{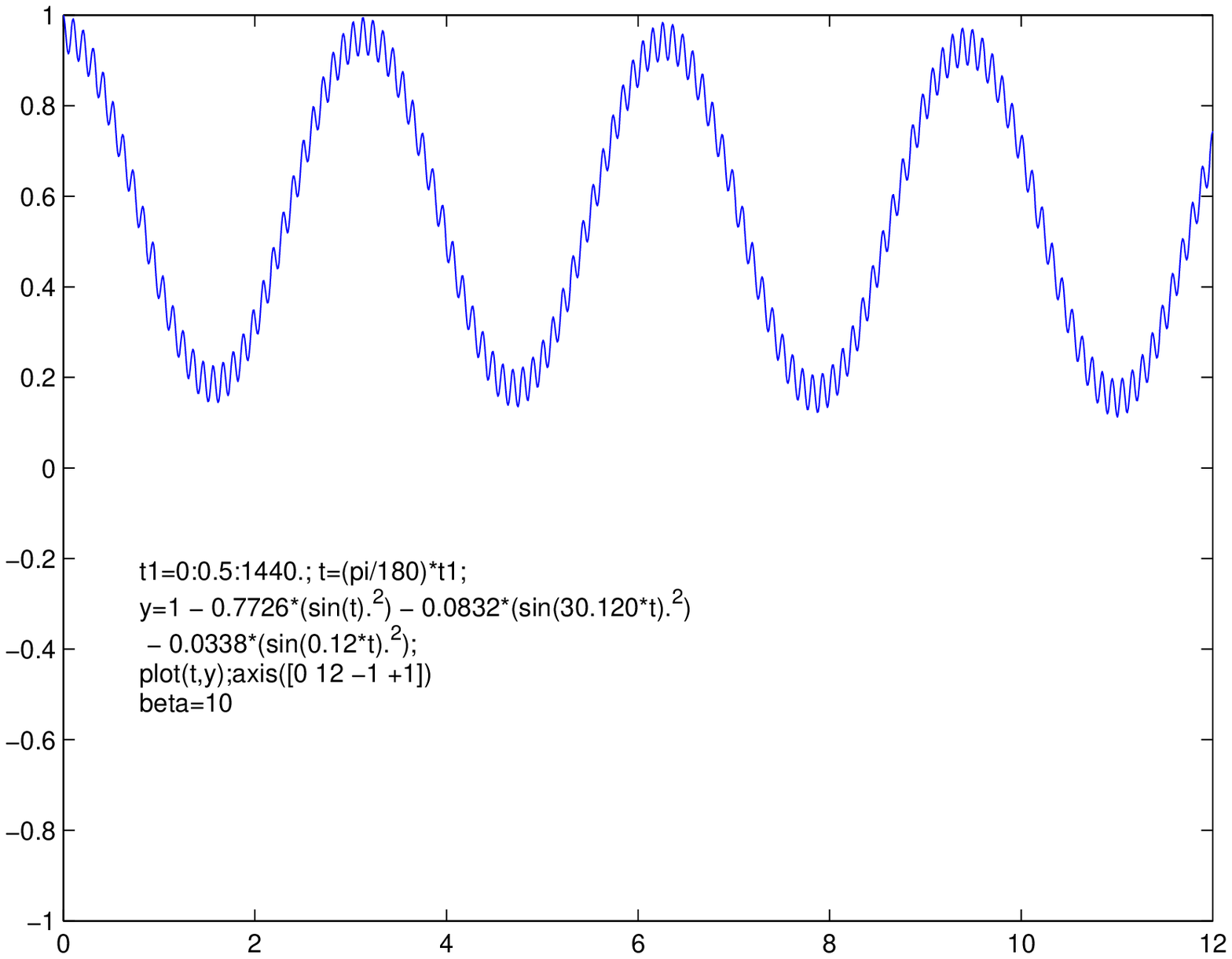}
   \end{center}
   \end{figure}

{\sf Figure~6:}~ $P(\nu_e \to \nu_e)(t)$ at $\theta = 32.45^o,
\beta
= 10^o, (L_{1 3}/L_{1 2} = 0.12)$. \\

\begin{figure}[h!]
   \begin{center}
   \includegraphics[width=10cm]{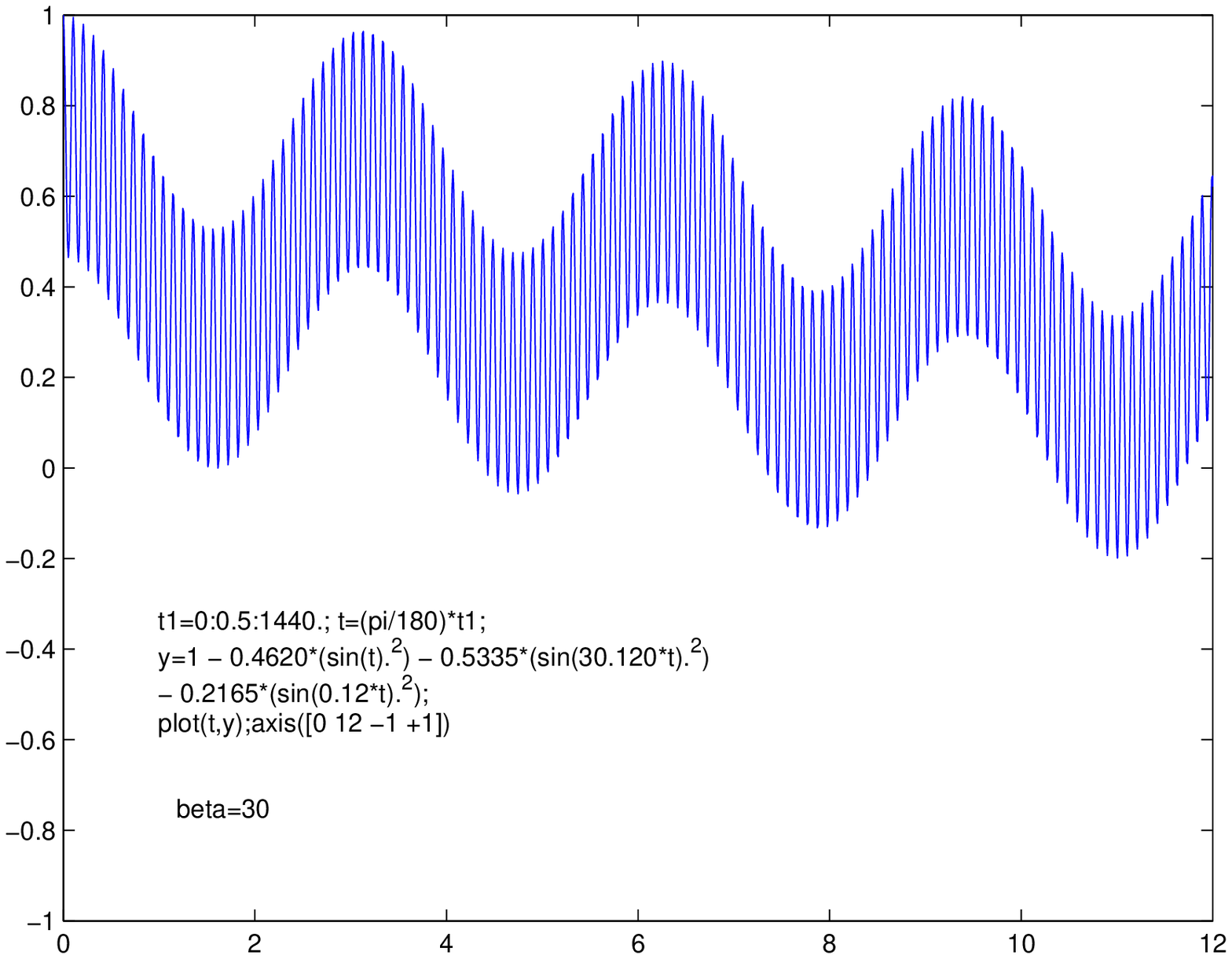}
   \end{center}
   \end{figure}

   {\sf Figure~7:}~ $P(\nu_e \to \nu_e)(t)$ at $\theta = 32.45^o, \beta
= 30^o, (L_{1 3}/L_{1 2} = 0.12)$. \\

\newpage
\begin{figure}[h!]
   \begin{center}
   \includegraphics[width=10cm]{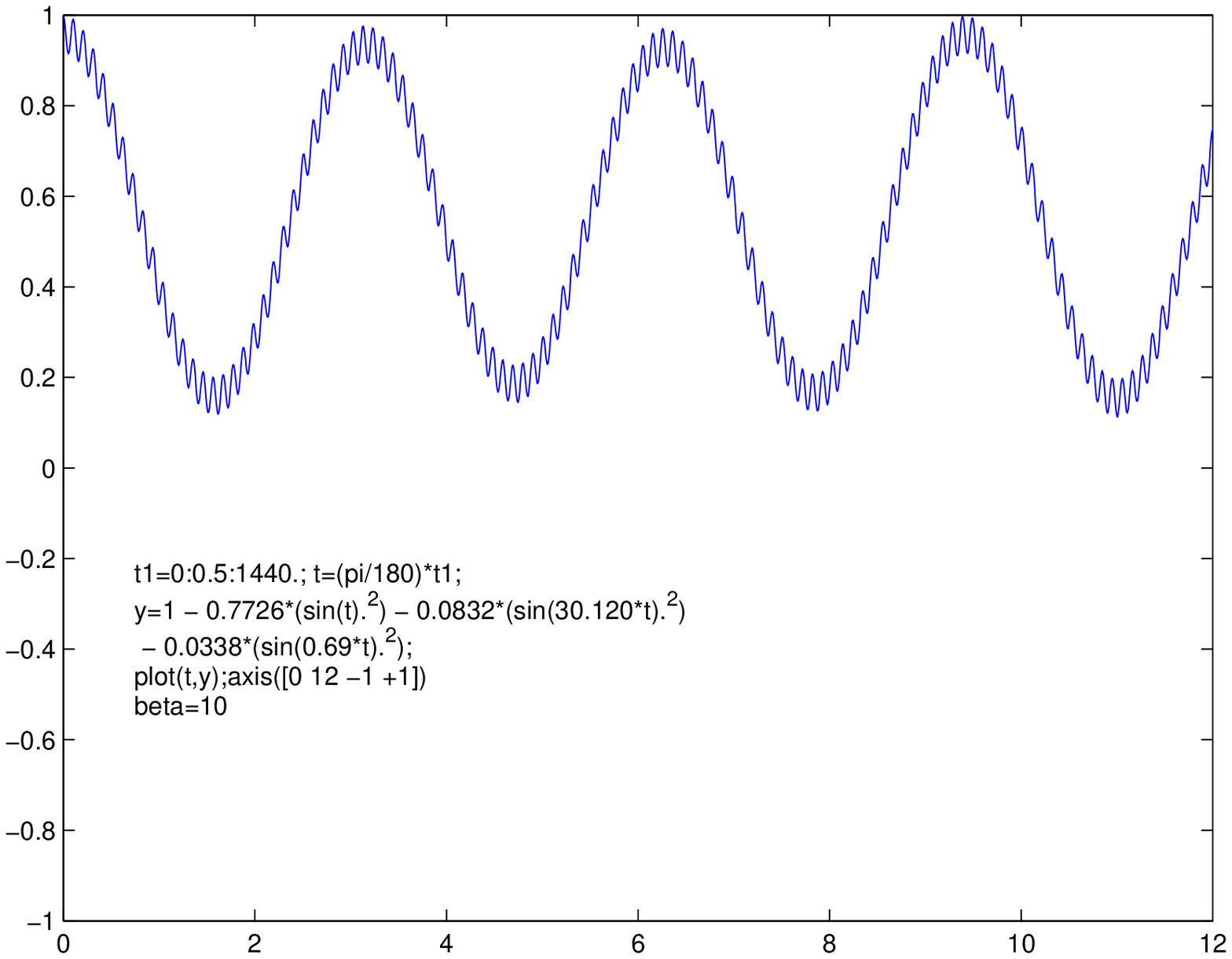}
   \end{center}
   \end{figure}

{\sf Figure~8:}~ $P(\nu_e \to \nu_e)(t)$ at $\theta = 32.45^o,
\beta = 10^o, (L_{1 3}/L_{1 2} = 0.69)$. \\

\begin{figure}[h!]
   \begin{center}
   \includegraphics[width=10cm]{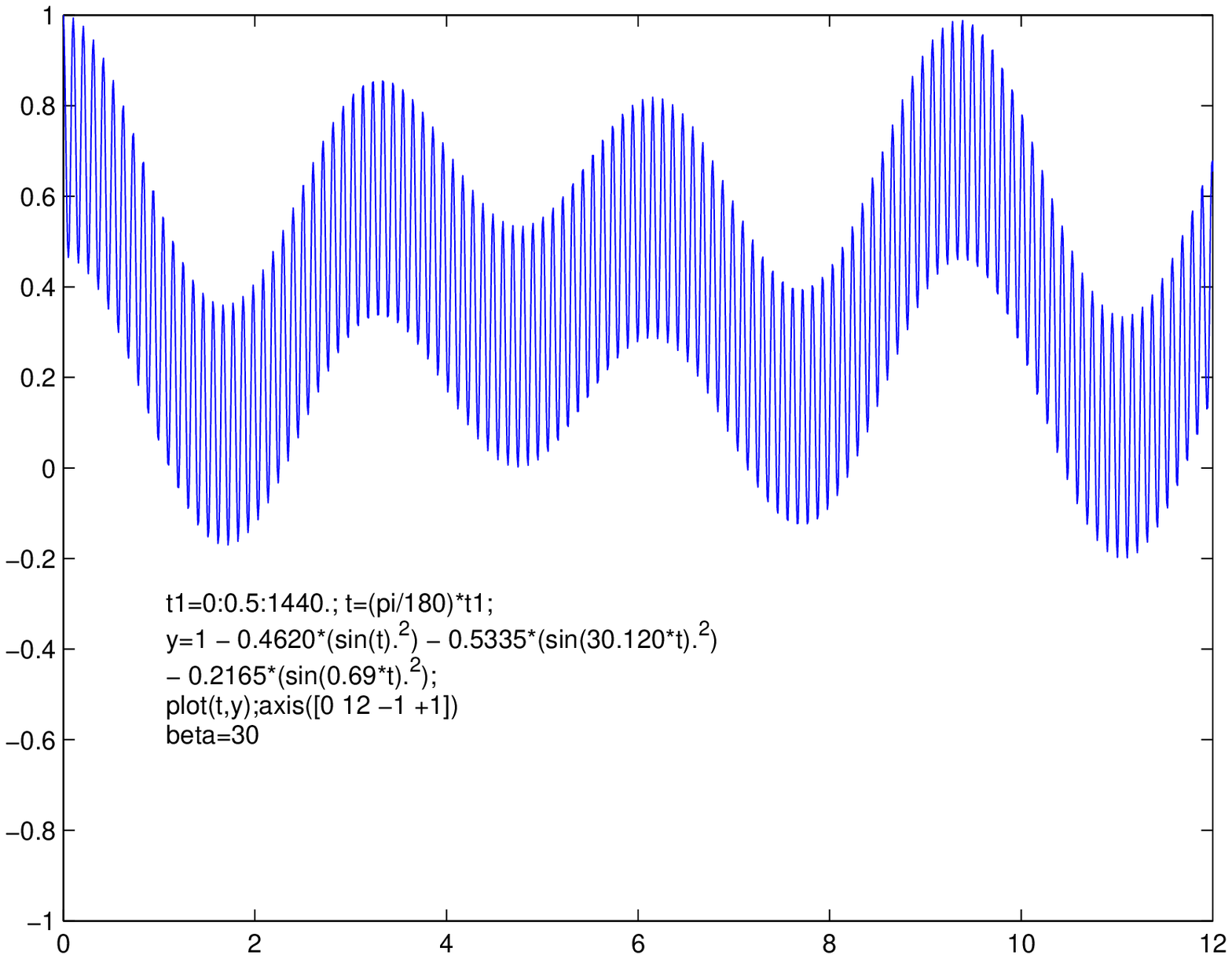}
   \end{center}
   \end{figure}

   {\sf Figure~9:}~ $P(\nu_e \to \nu_e)(t)$ at $\theta = 32.45^o, \beta
= 30^o, (L_{1 3}/L_{1 2} = 0.69)$. \\

\newpage
\begin{figure}[h!]
   \begin{center}
   \includegraphics[width=10cm]{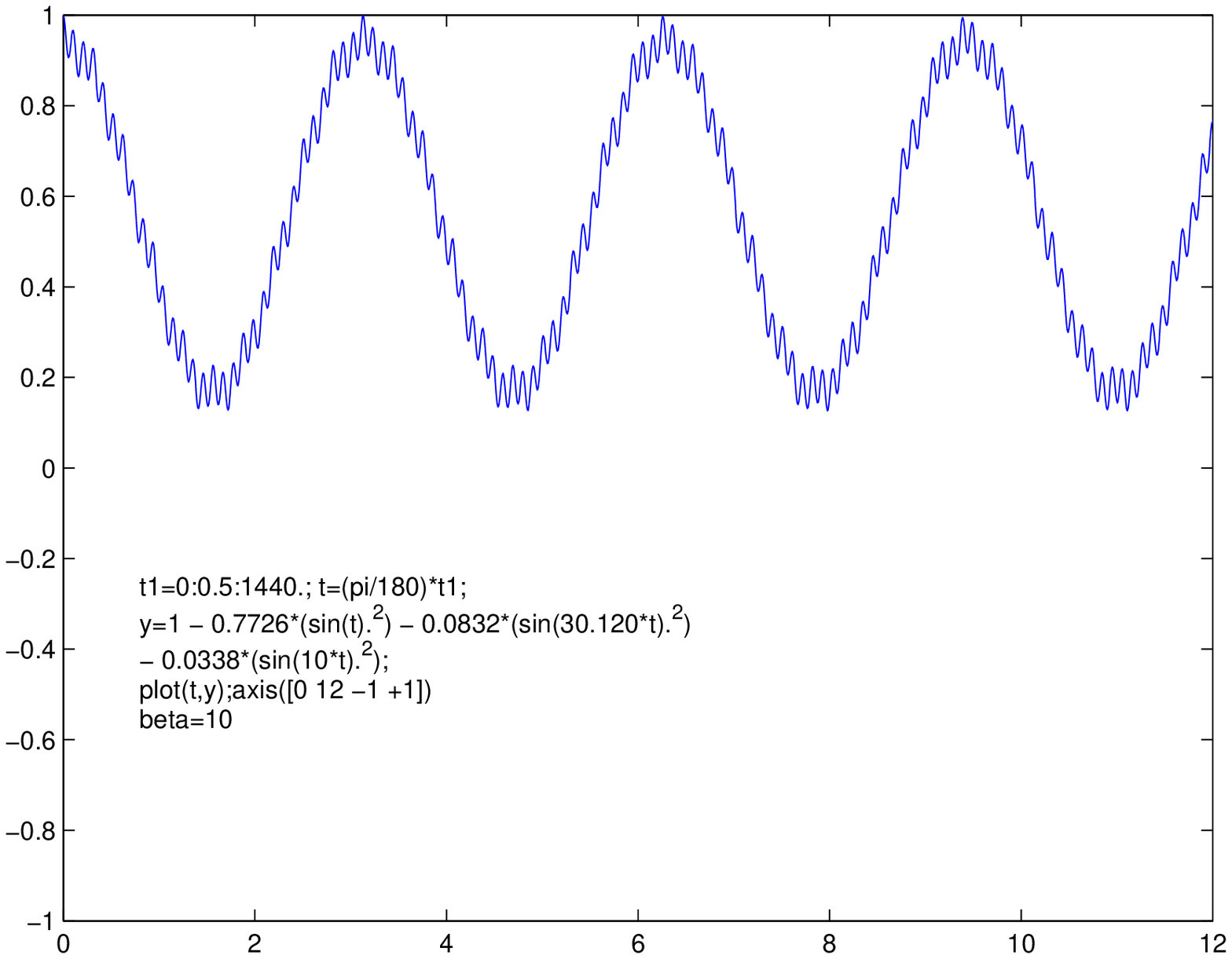}
   \end{center}
   \end{figure}

{\sf Figure~10:}~ $P(\nu_e \to \nu_e)(t)$ at $\theta = 32.45^o,
\beta = 10^o, (L_{1 3}/L_{1 2} = 10)$. \\

\begin{figure}[h!]
   \begin{center}
   \includegraphics[width=10cm]{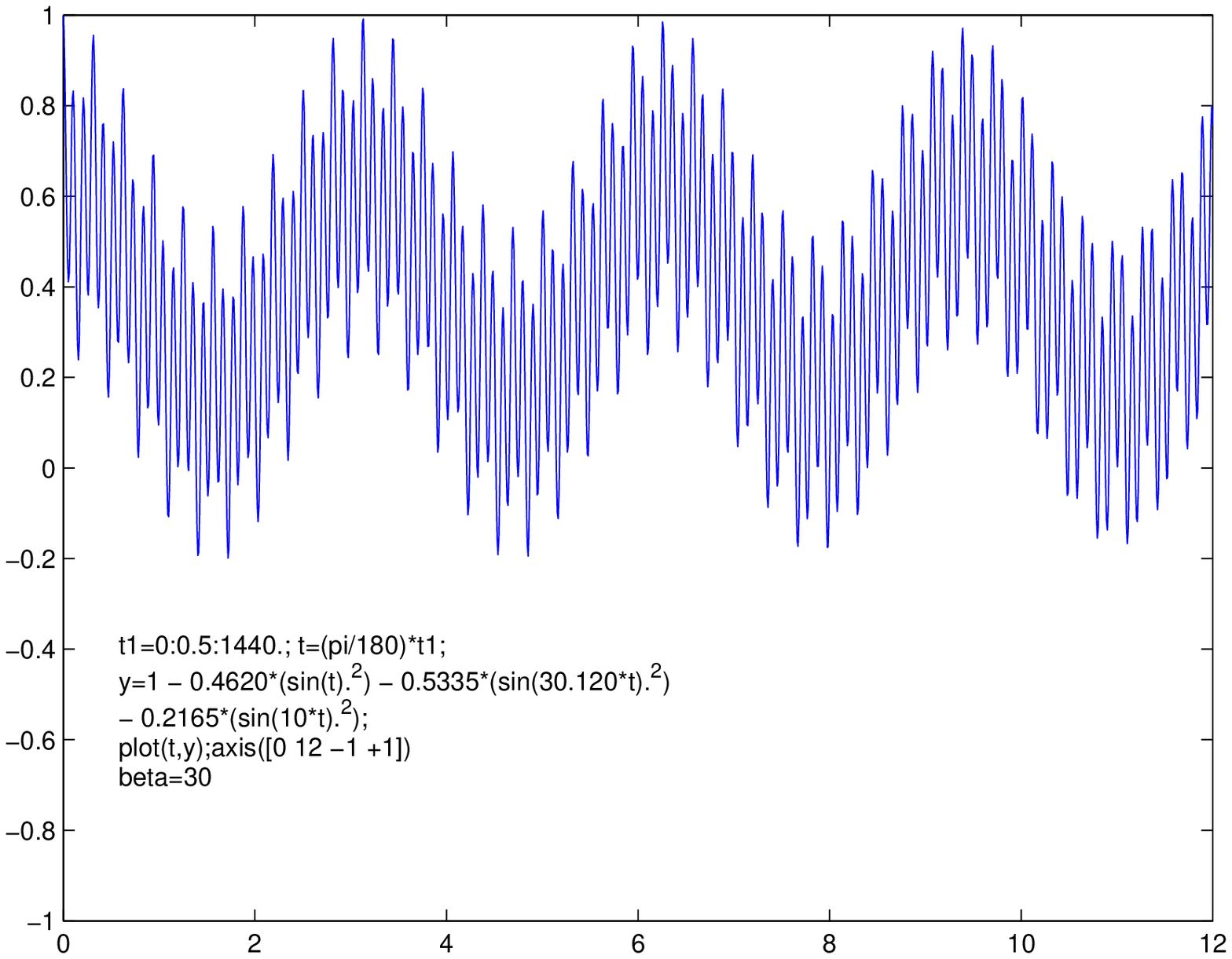}
   \end{center}
   \end{figure}

   {\sf Figure~11:}~ $P(\nu_e \to \nu_e)(t)$ at $\theta = 32.45^o, \beta
= 30^o, (L_{1 3}/L_{1 2} = 10)$. \\

\par
So, we see that at strict fulfillment of condition $\Delta m^2_{1
3} = \Delta m^2_{1 2} + \Delta m^2_{2 3}$ the expression for
probability of $\nu_e \to \nu_e$ transitions $P(\nu_e \to
\nu_e)(t)$ is positively defined while at any deviation from this
condition in order to make this expression for probability
positively defined, it is necessary to put a limitation on angle
mixing $\beta$ (i.e. the value for $\beta$ must be $\beta \le 15^o
\div 17^o$).

\section{Conclusion}

\par
The numeral value of sum $Sum$ of coefficients in expression (9)
connected with mixings
$$
Sum = cos^4(\beta)sin^2(2 \theta) + cos^2(\theta) sin^2(2 \beta) +
sin^2(\theta) sin^2(2 \beta)
$$
is larger than one (for example for $\theta = 32.45^o, \beta =
45^o, Sum = 1.2053$). In order to do the values of $P_{\nu_e \to
\nu_e}(t)$ positively defined, the maxima of oscillation
components of this expression don't have to coincide. Examination
has shown that for all the considered cases this condition is
fulfilled.
\par
This work has shown that at strict fulfillment of condition
$\Delta m^2_{1 3} = \Delta m^2_{1 2} + \Delta m^2_{2 3}$ the
expression for probability of $\nu_e \to \nu_e$ transitions
$P_{\nu_e \to \nu_e}(t)$ is positively defined at every values of
$\theta$ and $\beta$ while at any arbitrarily small deviation from
this condition it becomes negative and then in order to make this
expression for probability positively defined it is necessary to
put a limitation on angle mixing $\beta$ at fixed value of $\theta
= 32.45^o$ (i.e. the value for $\beta$ must be $\beta \le 15^o
\div 17^o$).

\newpage
\par
{\bf References}\\

\par
\noindent 1. B. M. Pontecorvo , Soviet Journ. JETP, 33 (1957) 549;
\par
JETP, 34 (1958) 247.
\par
\noindent 2. Z. Maki et al., Prog.Theor. Phys., 28 (1962) 870.
\par
\noindent 3. B. M. Pontecorvo, Soviet Journ. JETP, 53 (1967) 1717.
\par
\noindent 4. Kh. M. Beshtoev, JINR Communication E2-2004-58,
Dubna, 2004;
\par
hep-ph/0406124; JINR Communication E2-2005-123, Dubna, 2005,
\par
hep-ph/0506248;
\par
\noindent 5. S.M. Bilenky, B.M. Pontecorvo, Phys. Rep., C41 (1978)
225;
\par
F. Boehm, P. Vogel, Physics of Massive Neutrinos: Cambridge
\par
Univ. Press, 1987, p.27, p.121;
\par
S. M. Bilenky, S. P. Petcov, Rev. of Mod.  Phys., 59 (1977) 631.
\par
 V. Gribov, B. M. Pontecorvo, Phys. Lett. B, 28 (1969) 493.
\par
\noindent 6. L. Maiani, Proc. Intern. Symp.  on Lepton--Photon
Interaction,
\par
DESY, Hamburg., 1077, P.867.
\par
\noindent 7. K. Eguchi et al., Phys. Rev. Let. 90 (2003) 021802;
\par
T. Mitsui, 28-th Intern. Cosmic Ray Conf., Japan, 1 (2003) 1221;
\par
G. Gtatta, Report on the XXIst Inter. Conf. on Neutrino Physics
and
\par
Astrophysics, Paris, France, June 2004.
\par
\noindent 8. A. Habig, Proceedings of Inter. Cosmic Ray Conf.,
Japan, 1 (20030)
\par
1255;
\par
Ed. Kearns, Super-Kamiokande Collaboration, Report on Intern.
\par
Conf. Neutrino 2004, Paris, 2004.
\par
\noindent 9. Kh. M. Beshtoev, JINR Communication E2-2006-16,
Dubna, 2006;
\par
hep-ph/0508122, 2005.

\end{document}